\documentclass[10pt,twocolumn,aps,prb,superscriptaddress]{revtex4-2}
\usepackage{graphicx,epstopdf,amsmath,amssymb,multirow,array,CJK,color}
\graphicspath{{./}{PDF/}{../PDF/}}
\usepackage[german,english]{babel}
\usepackage[colorlinks, linkcolor=blue,anchorcolor=blue,citecolor=blue,urlcolor=blue]{hyperref}
\usepackage{booktabs}
\usepackage{ragged2e}
\newcommand{\manualfigcaption}[2]{%
  \refstepcounter{figure}\label{#1}%
  \par\vspace{4pt}\noindent\parbox{\linewidth}{\small\setlength{\parindent}{0pt}\justifying\noindent Figure~\thefigure. #2}%
}
\newcommand{\manualtabcaption}[2]{%
  \refstepcounter{table}\label{#1}%
  \par\noindent\parbox{\linewidth}{\small\setlength{\parindent}{0pt}\justifying\noindent Table~\thetable. #2}%
  \par\vspace{3pt}%
}
\usepackage{bm}
\DeclareMathSizes{8.4}{8.4}{6}{5}

\begin{document}

\title{Symmetry-protected triplet Weyl complexes}

\author{Yun-Yun Bai}\email{These two authors contributed equally to this work.}
\affiliation{State Key Laboratory of Metastable Materials Science and Technology $\&$ Hebei Key Laboratory of Microstructural Material Physics, School of Science, Yanshan University, Qinhuangdao 066004, China}

\author{Ke-Xin Pang}\email{These two authors contributed equally to this work.}
\affiliation{State Key Laboratory of Metastable Materials Science and Technology $\&$ Hebei Key Laboratory of Microstructural Material Physics, School of Science, Yanshan University, Qinhuangdao 066004, China}

\author{Yan Gao}\email{yangao9419@ysu.edu.cn}
\affiliation{State Key Laboratory of Metastable Materials Science and Technology $\&$ Hebei Key Laboratory of Microstructural Material Physics, School of Science, Yanshan University, Qinhuangdao 066004, China}

\author{Weikang Wu}\email{weikang\_wu@sdu.edu.cn}
\affiliation{Shenzhen Research Institute of Shandong University, Shenzhen, Guangdong 518057, China}

\author{Shengyuan A. Yang}
\email{shengyuan.yang@polyu.edu.hk}
\affiliation{Research Laboratory for Quantum Materials, Department of Physics and Materials, The Hong Kong Polytechnic University, Kowloon, Hong Kong, China}


\begin{abstract}
The Nielsen-Ninomiya theorem dictates that Weyl nodes must appear in pairs of opposite chirality to preserve global charge neutrality. However, in crystals, specific crystalline symmetries can stabilize multi-Weyl nodes, circumventing this pairwise constraint and enabling compensated Weyl complexes with mixed chiral charges.
The minimal configuration of this type is a triplet Weyl complex (TWC), comprising exactly three Weyl nodes.
Here, we systematically investigate the symmetry conditions required to realize TWCs. By screening all 1651 magnetic space groups (MSGs) in both spinless and spinful systems, we establish that: (i) Only TWCs with charge magnitudes of $\{1,1,2\}$ and $\{1,2,3\}$ are permitted; (ii) the $\{1,1,2\}$ configuration can be realized in 166 spinless MSGs and 70 spinful MSGs; and (iii)
the $\{1,2,3\}$-TWCs, which has not been reported before, can occur in 10 MSGs for both spinless and spinful cases.
We explicitly demonstrate the existence of $\{1,2,3\}$-TWC in a tight-binding model. Furthermore, we present the first electronic realization of $\{1,1,2\}$-TWC topological semimetal state in the chiral carbon allotrope DZQH-C$_{36}$, in which the three Weyl nodes form a collinear configuration, leading to a characteristic ``S''-shaped surface Fermi arc pattern.
Our findings uncover novel topological states featuring mixed chiral charges and provide guidance for exploring their physics in concrete material systems.

\end{abstract}

\maketitle


The Nielsen-Ninomiya no-go theorem~\cite{nogo1,nogo2}, a fundamental result in lattice field theory, mandates that Weyl nodes in a lattice must appear in pairs of opposite chirality.
In the past decade, the realization of Weyl nodes in the electronic band structure of so-called Weyl semimetals has attracted great interest~\cite{Armitage,WanXG,BurkovAA,TSMs}. In this solid-state context, it is noted that the influence of crystalline symmetry may help to circumvent the pairwise constraint imposed by Nielsen-Ninomiya theorem, while still preserving the global chiral charge neutrality. For example,
Ref.~\cite{Cno-go} proposed a configuration with a single Weyl node enclosed by a symmetry-enforced nodal wall~\cite{NSSMs}, where the wall carries a chiral charge opposite to the node. Another possibility is to merge multiple chiral charges into a multi-Weyl node~\cite{FangC,YangGF,HeQiu,ZhangT,XuG,HuangSM,WangHourglass,ChenXiao,LiuXu,ZhangC4,ChenC4,LiuC4}, so that one may have a Weyl complex comprising an odd number of Weyl nodes and having mixed chiral charges~\cite{WangXu,C-2-WPhonos,TTWC-II}.


Evidently, a triplet Weyl complex (TWC), consisting of exactly three Weyl nodes, represents the minimal such configuration.
Previous studies reported only a few particular TWC examples, all having chiral charge magnitudes of $\{1, 1, 2\}$~\cite{WangXu,C-2-WPhonos}.
Several fundamental questions about TWCs remain unexplored. These include: What are all possible combinations of chiral charges in a TWC? What is the distribution of the three Weyl nodes in Brillouin zone (BZ)? And what are the symmetry conditions allowing each TWC configuration? In addition, noting that the existing reports of TWC are all limited to phononic spectra~\cite{WangXu,C-2-WPhonos}, it remains an important task to find a fermionic realization of TWC in an electronic band structure.


\begin{figure*}[!t]
		\centering
		\includegraphics[width=0.98\textwidth]{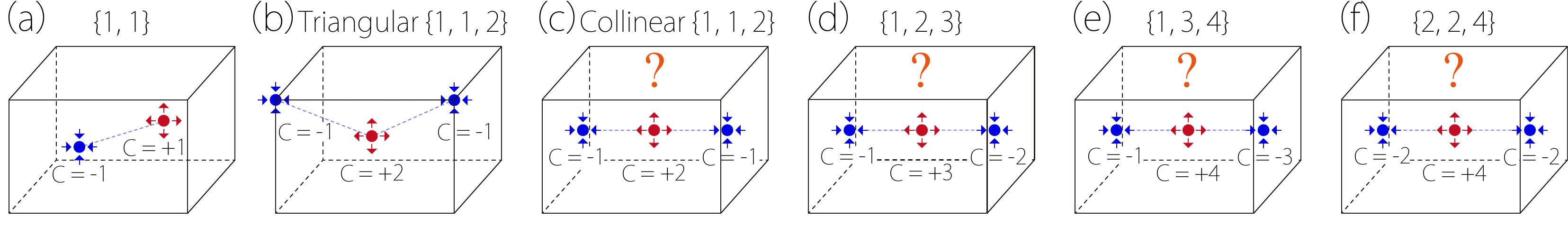}
		\manualfigcaption{fig_diagram}{(a) Conventional Weyl pair with topological charges $C = \pm 1$. (b-f) TWCs with mixed charges. (b) $\{1,1,2\}$-TWC with triangular configurations has been reported before. Here, we discover that (c)
$\{1,1,2\}$-TWC with collinear configuration and (d) $\{1,2,3\}$-TWC can exist in crystals. Meanwhile, (e) $\{1,3,4\}$-TWC and (f) $\{2,2,4\}$-TWC cannot be stabilized.}
\end{figure*}

We address these open questions in this work. By performing a comprehensive symmetry analysis of all 1651
magnetic space groups (MSGs) without and with spin-orbit coupling, we discover the following. (i) There are only two possible combinations of chiral charge magnitudes in a TWC: $\{1, 1, 2\}$ and $\{1, 2, 3\}$. (ii)
The $\{1, 1, 2\}$-TWC can be realized in 166 spinless MSGs and 70 spinful MSGs. Interestingly, besides the previously reported cases where the three Weyl nodes form a triangular configuration in BZ [see Fig.~\ref{fig_diagram}(b)], we identify a new configuration where the Weyl nodes have a collinear distribution [see Fig.~\ref{fig_diagram}(c)], all sitting on a high-symmetry axis. (iii) The
$\{1, 2, 3\}$-TWC [Fig.~\ref{fig_diagram}(d)], which was unknown before, can exist in 10 MSGs for both spinless and spinful cases. Here, the three Weyl nodes must have a collinear distribution on a sixfold screw axis.
Guided by the obtained symmetry conditions, we explicitly demonstrate the realization of $\{1, 2, 3\}$-TWC in a lattice model. Furthermore, we identify the collinear $\{1, 1, 2\}$-TWC in the electronic band structure of a chiral carbon allotrope DZQH-C$_{36}$. Using first-principles calculations, we show that this material is an almost ideal TWC semimetal.
The collinear distribution of the three Weyl nodes along $k_z$ axis leads to a characteristic ``S''-shaped Fermi arc on the side surfaces. These findings reveal previously unknown topological states, settle fundamental questions regarding TWCs, and
offer guidance for exploring novel topological material platforms.




{\color{blue}\emph{Symmetry analysis.}}
Denote the chiral charges of Weyl nodes in a TWC as $C_i$ $(i=1,2,3)$.
It was shown that the magnitude $|C|$ of a chiral charge that can be stabilized in crystals is limited to 1, 2, 3, and 4~\cite{YuEncyclopedia,Liu-type-III-EP,Zhang-type-IV-EP}.

Under the requirements of exactly three nodes and global charge neutrality, i.e.,
\begin{eqnarray}\label{cnc}
  C_1+C_2+C_3=0,
\end{eqnarray}
it is easy to see that
there are only four possible candidates: $\{1, 1, 2\}$, $\{1, 2, 3\}$, $\{1, 3, 4\}$, and $\{2, 2, 4\}$.



Denote the locations of the three Weyl nodes as $\bm k_i$ $(i=1,2,3)$. One notes that an elementary Weyl node with $|C|=1$ can be sitting at any generic $k$-point in BZ, but a multi-Weyl node with $|C|>1$ has to be located at a high-symmetry $\bm k$ with a nontrivial little group $G_{\bm k}$. To realize a TWC by a given MSG $G$, the MSG must satisfy the following conditions.

\begin{enumerate}
\item The MSG $G$ must allow the coexistence of chiral charges $|C_1|$, $|C_2|$, and $|C_3|$.
\item To ensure there are exactly three Weyl nodes, for $\{1, 2, 3\}$ and $\{1, 3, 4\}$ TWCs,
each $k$-point $\bm k_i$ must remain invariant under all point group operations in $G$. In other words, the size of $\text{Star}(\bm k_i)$ must be one. Otherwise, there would be duplicated Weyl nodes required by symmetry.

\item For $\{1, 1, 2\}$ and $\{2, 2, 4\}$ TWCs, the two Weyl nodes with identical charge, denoted by $i=1$ and $i=2$, may or may not be connected by symmetry. For the former case, condition 2 applies. For the latter case,
    the size of $\text{Star}(\bm k_i)$ $(i=1,2)$ must be two, and the symmetry operation $O$ that connects $\bm k_1$ and $\bm k_2$ must not involve any improper rotation.

\item Actually, since a chiral charge must have a sign flip under an improper rotation (mirror or inversion), the point group of $G$ cannot contain any such operations.

\end{enumerate}

Under these conditions, we perform a comprehensive search over the 1651 MSGs,
based on the encyclopedia of emergent particles and compatibility relations~\cite{YuEncyclopedia,Liu-type-III-EP,Zhang-type-IV-EP,MTQC}, considering both the absence and presence of spin-orbit coupling. Our analysis reveals that the $\{1, 3, 4\}$ and $\{2, 2, 4\}$ TWCs cannot be stabilized in crystals.

\begin{figure}[!t]
	\centering
	\includegraphics[width=0.9\columnwidth]{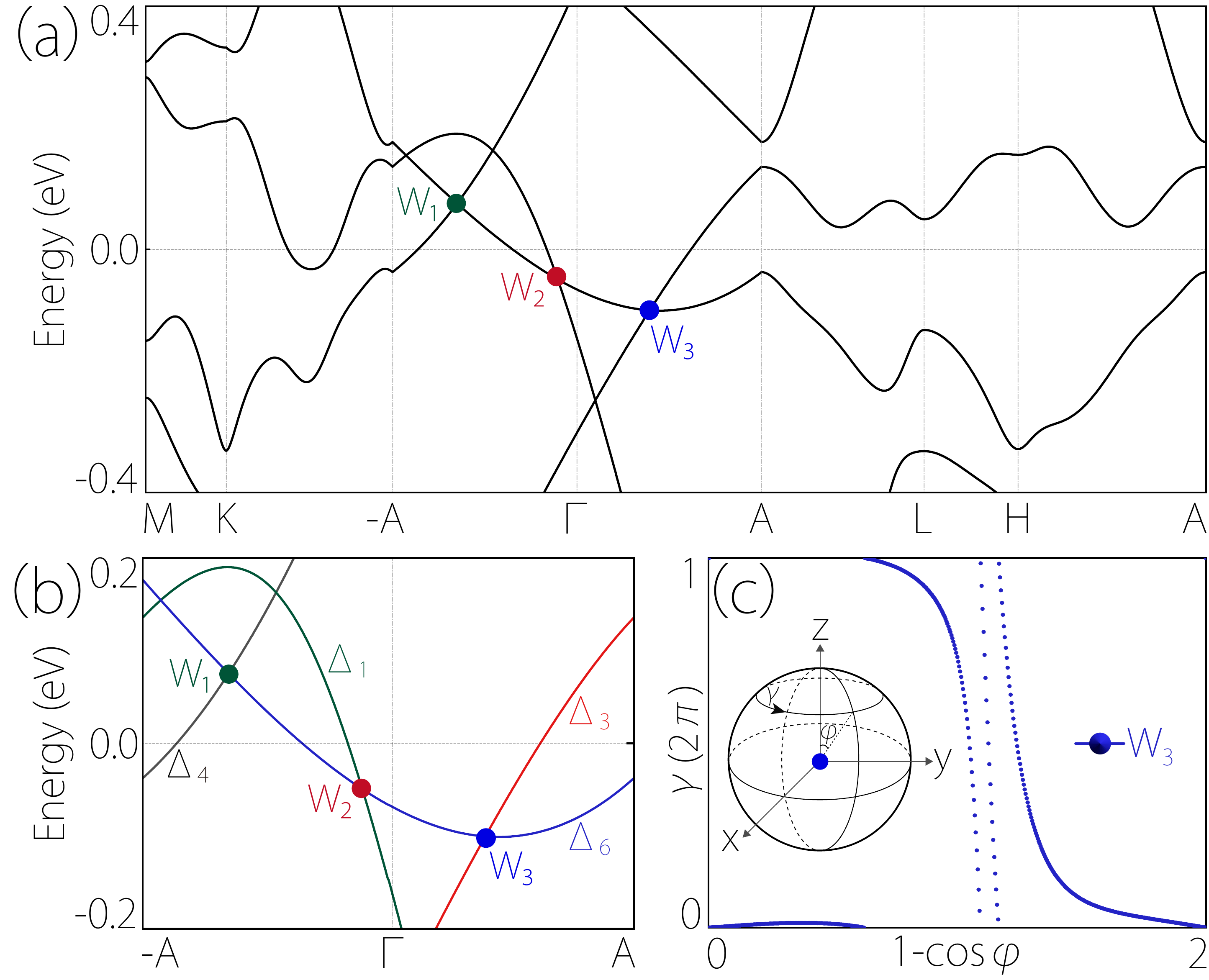}
	\manualfigcaption{fig_123}{(a) Band structure of our constructed $\{1,2,3\}$-TWC lattice model. The highlighted three crossing points $W_1$, $W_2$, and $W_3$ are the three Weyl nodes forming the TWC. (b) Enlarged band structure along -A-$\Gamma$-A path. The symmetry characters of the low-energy bands are labeled. (c) Evolution of Berry phase
on a latitude circle which sweeps through a small sphere enclosing $W_3$ (see the inset). The phase winding confirms $W_3$ is a cubic Weyl node.}
\end{figure}

The $\{1, 1, 2\}$-TWC can exist in 166 spinless MSGs and 70 spinful MSGs. The detailed results, including MSG elements, the distribution of Weyl nodes in BZ, and the symmetry characters of crossing bands, are presented in the Table SIII of the Supplemental Material (SM)~\cite{39SM}.
Notably, we find a new class of collinear $\{1, 1, 2\}$ TWCs, which is likewise symmetry allowed in these 166 spinless and 70 spinful MSGs.
Different from the previously reported triangular configuration, the three nodes in a collinear $\{1, 1, 2\}$ TWCs
are located on the same high-symmetry path in BZ [see Fig.~\ref{fig_diagram}(c)].

\begin{table*}[!t]
\noindent\begin{minipage}{\textwidth}
\manualtabcaption{tab:msg_123}{Magnetic space groups (MSGs) that support $\{1,2,3\}$-TWCs (in both the spinless and spinful cases). The symmetry generators and the distribution of the Weyl nodes in Brillouin zone are also presented.}
\centering
\begingroup
\fontsize{8.4}{9.6}\selectfont
\setlength{\tabcolsep}{1.5pt}
\renewcommand{\arraystretch}{1.00}
\renewcommand{\multirowsetup}{\centering}
\begin{tabular}{@{}>{\centering\arraybackslash}m{0.0948\textwidth}>{\centering\arraybackslash}m{0.1327\textwidth}>{\centering\arraybackslash}m{0.0986\textwidth}|>{\centering\arraybackslash}m{0.0948\textwidth}>{\centering\arraybackslash}m{0.2275\textwidth}>{\centering\arraybackslash}m{0.0986\textwidth}|>{\centering\arraybackslash}m{0.2130\textwidth}@{}}
\hline\hline
\multicolumn{3}{c|}{\rule[-0.16cm]{0pt}{0.46cm}Type-I MSGs} &
\multicolumn{3}{c|}{\rule[-0.16cm]{0pt}{0.46cm}Type-III MSGs} &
\multirow[c]{2}{0.2130\textwidth}{\centering Nodes distribution} \\
\cline{1-6}
\rule[-0.16cm]{0pt}{0.46cm}MSG & Generators & Location &
MSG & Generators & Location & \\
\hline
\rule[-0.28cm]{0pt}{0.70cm}169.113 & $\{C_{6}^{+}|00\tfrac{1}{6}\}$ &
&
178.159 & $\{C_{6}^{+}|00\tfrac{1}{6}\}$; $\{C_{21}^{\prime}\mathrm{T}|000\}$ &
&
\multirow[c]{5}{0.2130\textwidth}{\centering\includegraphics[width=0.88\linewidth,height=3.10cm,keepaspectratio]{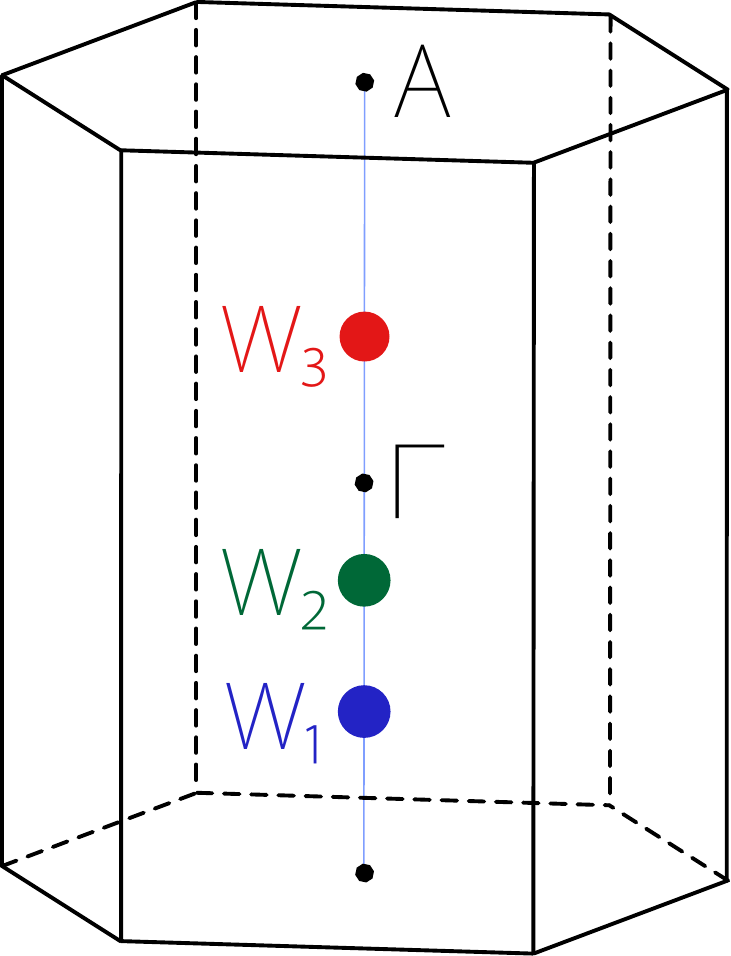}} \\
\rule[-0.28cm]{0pt}{0.70cm}170.117 & $\{C_{6}^{+}|00\tfrac{5}{6}\}$ & &
179.165 & $\{C_{6}^{+}|00\tfrac{5}{6}\}$; $\{C_{21}^{\prime}\mathrm{T}|000\}$ & & \\
\rule[-0.28cm]{0pt}{0.70cm}171.121 & $\{C_{6}^{+}|00\tfrac{1}{3}\}$ & $\Gamma$--A &
180.171 & $\{C_{6}^{+}|00\tfrac{1}{3}\}$; $\{C_{21}^{\prime}\mathrm{T}|000\}$ & $\Gamma$--A & \\
\rule[-0.28cm]{0pt}{0.70cm}172.125 & $\{C_{6}^{+}|00\tfrac{2}{3}\}$ & &
181.177 & $\{C_{6}^{+}|00\tfrac{2}{3}\}$; $\{C_{21}^{\prime}\mathrm{T}|000\}$ & & \\
\rule[-0.28cm]{0pt}{0.70cm}173.129 & $\{C_{6}^{+}|00\tfrac{1}{2}\}$ & &
182.183 & $\{C_{6}^{+}|00\tfrac{1}{2}\}$; $\{C_{21}^{\prime}\mathrm{T}|000\}$ & & \\
\hline\hline
\end{tabular}
\endgroup
\end{minipage}
\end{table*}

More importantly, we discover a previously unknown $\{1, 2, 3\}$-TWC, which can be stabilized in 10 MSGs both without and with spin-orbit coupling. These ten MSGs are listed in Table~\ref{tab:msg_123}, five are type-I MSGs and the other five are of type-III, indicating that such systems must have magnetic ordering. One can also see that for all ten MSGs, there is a $C_6$ screw rotation symmetry (with different fractional translations), and the three Weyl nodes are collinearly distributed on the $k_z$-axis ($\Gamma$-$A$ path), which is invariant under $C_6$ rotation. More details are given in the SM~\cite{39SM}.

{\color{blue}\emph{$\{1,2,3\}$-TWC in lattice model.}}
To confirm the existence of the novel class of $\{1,2,3\}$-TWC, we explicitly demonstrate it in a lattice model.
The 10 MSGs in Table~\ref{tab:msg_123} all possess a sixfold screw axis, indicating that the number of bands in such a model has to be a multiple of six. Here, we construct a twelve-band spinless tight-binding model on a hexagonal lattice, constrained by the
type-I MSG 169.113 ($P6_1$). The details of this model are presented in the SM~\cite{39SM}.

Figure~\ref{fig_123} shows the low-energy band structure of this model. Near the Fermi level, the highest valence band and lowest conduction band form three linear crossings along the $k_z$ axis ($\Gamma$--A path), which are labeled as $W_i$ ($i=1,2,3$) in Fig.~\ref{fig_123}(a).

These band crossing points are symmetry-allowed, which can be seen by noting that the two crossing bands at each $W_i$ have different symmetry characters. To demonstrate this, in Fig.~\ref{fig_123}(b), we label the irreducible representation $\Delta_p$ of each band. Here, the $\{C_{6z}^+ | 00\frac{1}{6}\}$ eigenvalue of a band state at $k_z$ with $\Delta_p$ symmetry is $\lambda_p=\exp[i\pi (c_p-u)/3]$, where $c_p=(p-2)+\frac{2}{\sqrt{3}}\cos[\frac{2\pi(p-1)}{3}+\frac{\pi}{6}]$ and $u=c k_z / \pi$.
According to Ref.~\cite{FangC}, the magnitude of chiral charge $C$ of a Weyl node at the crossing between $\Delta_p$ and $\Delta_{p'}$ bands is determined by the ratio $\lambda_p/\lambda_{p'}$. Specifically,
\begin{equation}
  |C|=\frac{3}{\pi}\left|\ln(\lambda_p/\lambda_{p'})\right|.
\end{equation}

From Fig.~\ref{fig_123}(b), one can easily see that $W_3$ is a cubic Weyl node, since $\lambda_6/\lambda_{3}=e^{-i\pi}=-1$. In Fig.~\ref{fig_123}(c), we also directly evaluate the chiral charge of $W_3$, by tracing the evolution of Berry phase on a loop as it sweeps through a small sphere enclosing $W_3$. The result confirms $|C|=3$.

Electronic states around this cubic node $W_3$ can be studied by constructing the symmetry-constrained effective $k\cdot p$ model~\cite{Mkp}. We find that this model takes the form of
\begin{equation}
	\mathcal{H}_{W_3}(\bm{q}) = \epsilon(\bm q) + \big[(\alpha q_+^3 + \beta q_-^3) \sigma_+ + \text{H.c.}\big] + m(\bm{q}) \sigma_z,
\end{equation}
where energy and momentum $\bm q$ are measured from $W_3$, $\epsilon(\bm{q}) = w_1 q_z + w_2(q_x^2 + q_y^2)$, $m(\bm{q}) = m_1 q_z + m_2 (q_x^2 + q_y^2)$, $q_\pm=q_x\pm iq_y$, $\sigma_\pm=\sigma_x\pm i\sigma_y$, $\sigma$'s are Pauli matrices denoting the two crossing bands, $w_i$ and $m_i$ are real parameters, while $\alpha$ and $\beta$ are complex model parameters. 
The $q^3$-type splitting in the $q_x$-$q_y$ plane is a signature of cubic Weyl node having
$|C| = 3$. 

Similarly, one finds $W_2$ is a quadratic Weyl node ($|C|$=2), and $W_1$ is an elementary Weyl node ($|C|$=1), as also supported by their effective models (see \cite{39SM}). This confirms the existence of our claimed $\{1,2,3\}$-TWC.

We have two remarks here. First, as discussed above, the three nodes in a $\{1,2,3\}$-TWC must be arranged collinearly on the $k_z$-axis (see Table~\ref{tab:msg_123}).
Nevertheless, their sequence on the axis is not fixed, but depends on specific details of the system.

Second, under the charge neutrality condition Eq.~(\ref{cnc}), the chiral changes of $W_1$ and $W_2$ must have the same sign, opposite to that of $W_3$. Nevertheless, their specific signs, i.e., being $\{1,2,-3\}$ or $\{-1,-2,3\}$, depend on the model details. And by a mirror operation acting on the model, one can flip the signs of the charges.

{\color{blue}\emph{Collinear $\{1,1,2\}$-TWC in a carbon allotrope.}}
As mentioned, collinear $\{1,1,2\}$-TWC is also a new finding of this work. We identify an electronic realization of
such $\{1,1,2\}$-TWC in a chiral carbon allotrope DZQH-C$_{36}$.

The structure of DZQH-C$_{36}$ is shown in Fig.~\ref{fig_structure}. It consists of two basic units: a $sp^2$-hybridized distorted zigzag (DZ) nanoribbon wrapped into a triangular pillar [Fig.~\ref{fig_structure}(d)] and a $sp^3$-hybridized quadrilateral helical (QH) nanoribbon [Fig.~\ref{fig_structure}(e)]. Both units are extended along the $c$ axis, and they are interconnected in the lateral direction forming a hexagonal lattice pattern in the top view [see Figs.~\ref{fig_structure}(a) and~\ref{fig_structure}(b)].
Figure~\ref{fig_structure}(f) indicates the primitive cell, which contains 36 carbon atoms. Evidently, the helical unit endows the lattice with a handedness. The structure in Figs.~\ref{fig_structure}(a) and~\ref{fig_structure}(b) corresponds to the left-handed one, denoted as $l$-DZQH-C$_{36}$ [Fig.~\ref{fig_structure}(a)], having space group $P6_5$ (No.~170). Its mirror image gives the right-handed partner, $r$-DZQH-C$_{36}$ [Fig.~\ref{fig_structure}(b)], having space group $P6_1$ (No.~169).
The detailed structural information is given in the Table SI of SM~\cite{39SM}.

\begin{figure*}[!t]
\noindent\begin{minipage}{\textwidth}
	\centering
	\includegraphics[width=0.9\textwidth]{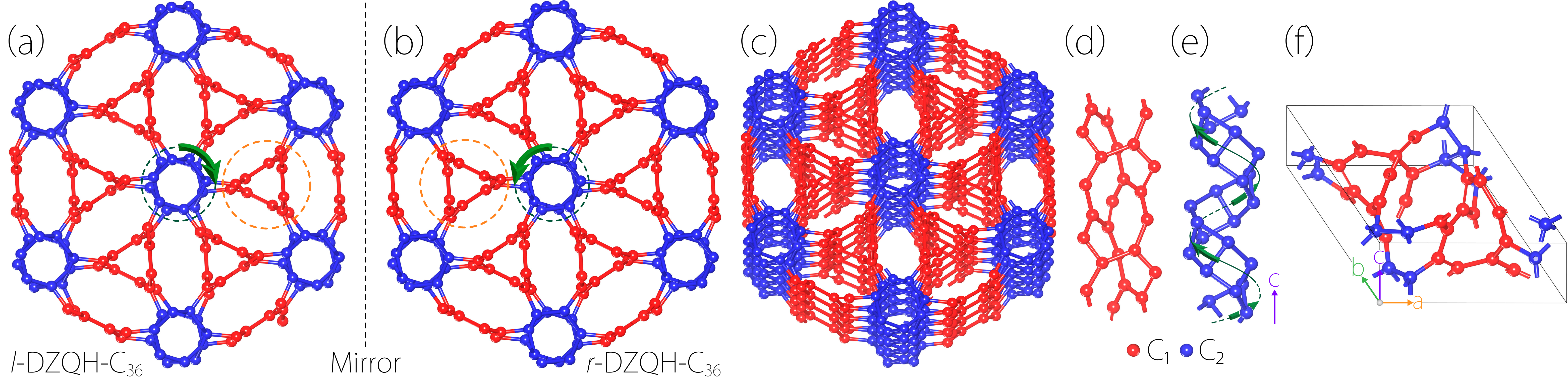}
	\manualfigcaption{fig_structure}{Crystal structure of DZQH-C$_{36}$. Top views of (a) left-handed and (b) right-handed DZQH-C$_{36}$, which are mirror images of each other. (c) Perspective view of $l$-DZQH-C$_{36}$. The structure consists of two basic building blocks, which are  (d) $sp^2$-hybridized distorted zigzag nanoribbon (red colored) and (e) $sp^3$-hybridized distorted quadrilateral helical nanoribbon (blue colored). (f) Primitive cell of $l$-DZQH-C$_{36}$.}
\end{minipage}
\end{figure*}

First-principles calculations indicate DZQH-C$_{36}$ enjoys excellent stability. Molecular dynamics simulation shows this allotrope can be stable up to 600 K [see Fig. S4 in the SM~\cite{39SM}]. Its formation energy ($\sim -8.22$ eV/atom) is comparable to, or even lower than, several known carbon allotropes, including the experimentally synthesized T-carbon ($-7.92$ eV/atom)~\cite{T-carbon-I,T-carbon-II}.

Importantly, the symmetry (MSG 169.114 or 170.118) of DZQH-C$_{36}$ belongs to one of the candidate groups that host collinear $\{1,1,2\}$-TWC in the spinless case (given the negligible spin-orbit coupling for carbon, the system can be safely treated as spinless). Figure~\ref{fig_band}(a) shows the calculated electronic band structure of $l$-DZQH-C$_{36}$.
One finds that DZQH-C$_{36}$ is a very good semimetal, with a dip in density of states at Fermi level.
The highest occupied band and lowest unoccupied band cross each other near Fermi level. One crossing point $W_2$ is at $\Gamma$ point (at energy $\sim-3.3$~meV, see the inset of ~\ref{fig_band}(a)), and another point $W_1$ is on $\Gamma$-$A$ path (at energy $\sim -19.2$ meV). In fact, there is a third crossing point $W_{1'}$ also on $\Gamma$-$A$, which is the time-reversal image of $W_1$ and is not shown in Fig.~\ref{fig_band}(a). Hence, there are three Weyl nodes on $\Gamma$-$A$, consistent with the collinear $\{1,1,2\}$-TWC that can be expected by symmetry [see Fig.~\ref{fig_band}(b)].

To confirm the character of $\{1,1,2\}$-TWC, we directly examine the chiral charges of the three nodes, by
numerically tracing the evolution of the Wannier charge center on a small sphere enclosing each node. The result shows
that $W_2$ has charge $+2$, while $W_1$ and $W_{1'}$ have charge $-1$, consistent with the $\{1,1,2\}$ configuration [Fig.~\ref{fig_band}(b)].
Chiral charges act as source and sink for Berry curvature field. In Fig.~\ref{fig_band}(c), we plot the distribution of Berry curvature field. One can clearly see that the field is emitted from $W_2$ node and absorbed by $W_1$ and $W_{1'}$ nodes.

The quadratic node at $\Gamma$ is protected by the sixfold screw rotation and time reversal symmetry. To characterize this node, we construct its $k\cdot p$ model from symmetry.
The result is
\begin{equation}\label{kp}
	\mathcal{H}_{W_2}(\bm k) = \epsilon(\bm{k}) + (\gamma k_+^2 \sigma_+ + \text{H.c.}) + v_z k_z \sigma_z
\end{equation}
where $\epsilon(\bm{k}) = \epsilon_1(k_x^2 + k_y^2) + \epsilon_2 k_z^2$, $k_\pm = k_x \pm i k_y$, $\epsilon_i$'s and $v_z$ are real parameters, and $\gamma$ is a complex model parameter. This model confirms $W_2$ is a quadratic Weyl node.

The discussion above is for left-handed DZQH-C$_{36}$. For $r$-DZQH-C$_{36}$, all the chiral charges have a sign flip [Fig.~\ref{fig_band}(d)], since the two are connected by a mirror operation, demonstrating the coupling between chiral charge and structural chirality.

\onecolumngrid
\vfill
\noindent\begin{minipage}{\textwidth}
	\centering
	\includegraphics[width=0.9\textwidth]{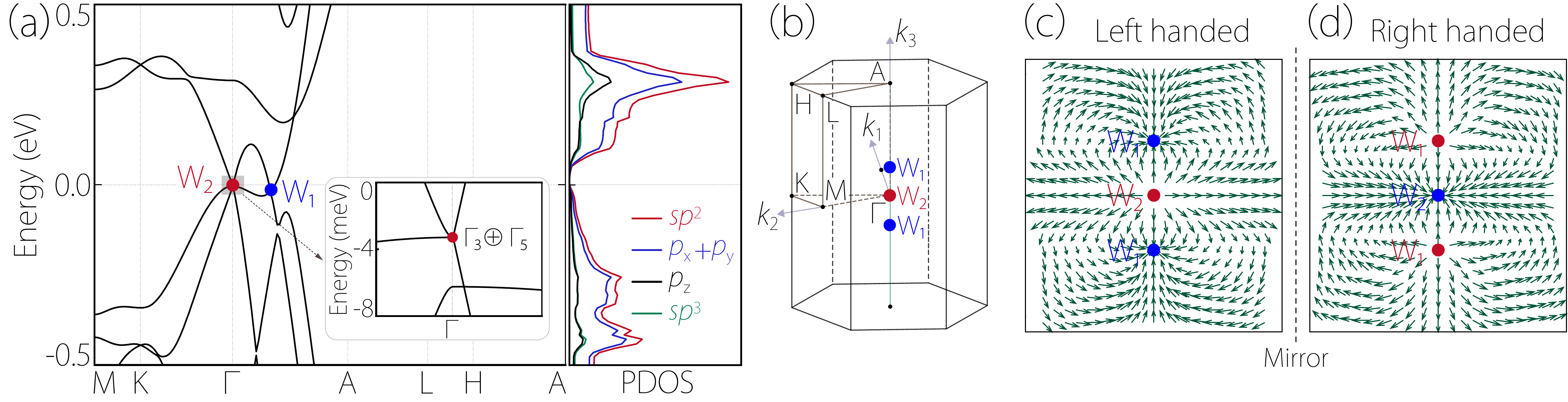}
	\manualfigcaption{fig_band}{(a) Calculated band structure of  $l$-DZQH-C$_{36}$, along with the projected density of states (PDOS). 
The inset enlarges the region around the quadratic Weyl node $W_2$. (b) Distribution of Weyl nodes in Brillouin zone, showing a $\{1,1,2\}$-TWC. (c,d) Distribution of Berry-curvature field in the $k_x$-$k_z$ plane for (c) $l$-DZQH-C$_{36}$ and (d) $r$-DZQH-C$_{36}$.}
\end{minipage}
\newpage
\twocolumngrid


\begin{figure}[!t]
	\centering
	\includegraphics[width=0.9\columnwidth]{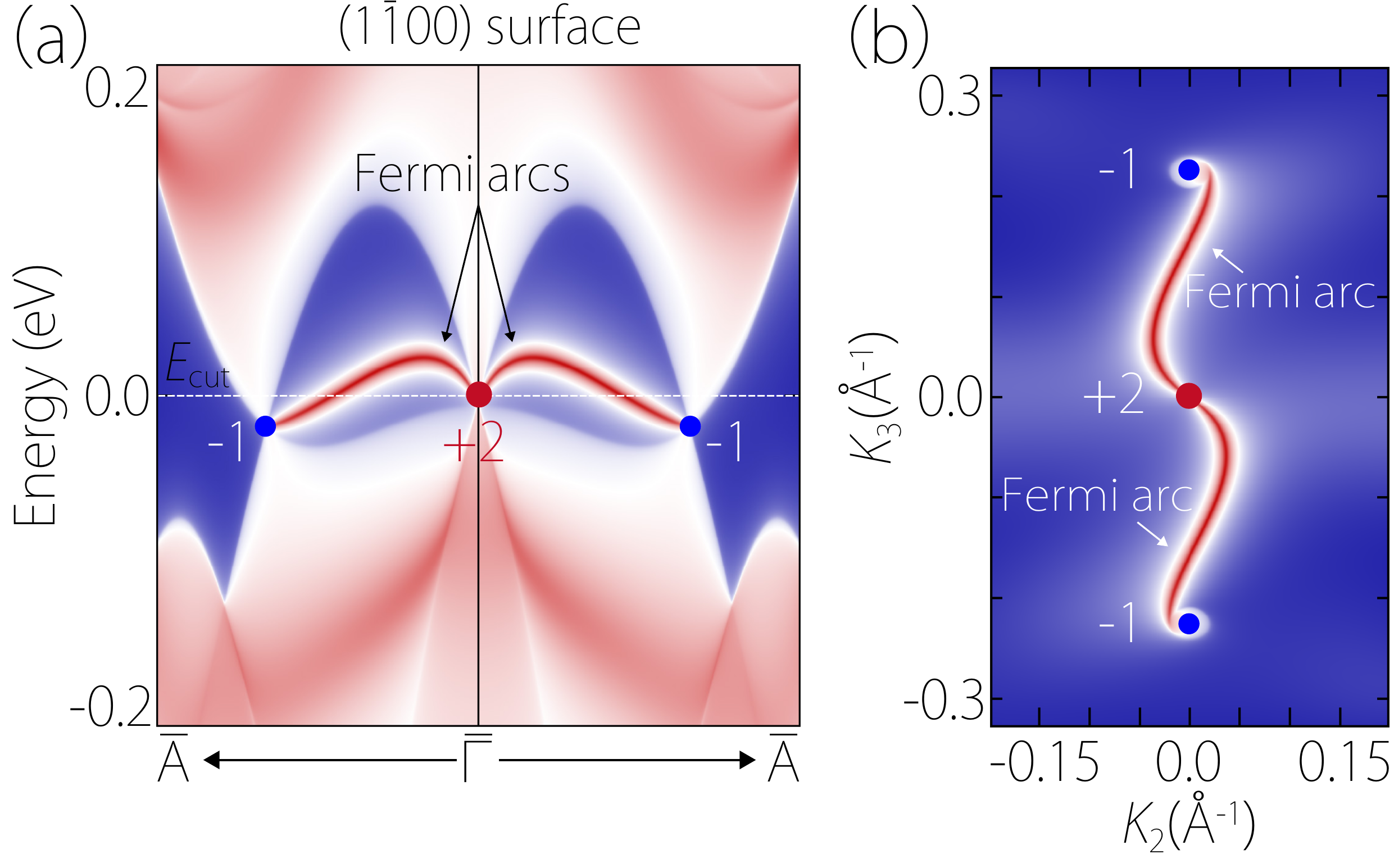}
	\manualfigcaption{fig_surface}{(a) Surface spectrum of $l$-DZQH-C$_{36}$ on the $(1\bar{1}00)$ surface. (b) Constant-energy slice at the Fermi level, showing characteristic ``S''-shaped Fermi arcs.}
\end{figure}

{\color{blue}\emph{Discussion.}}
We have clarified all possible TWCs in crystals and provided the corresponding symmetry conditions.
This unveils a new class of TWCs with $\{1, 2, 3\}$ chiral charge combination and a new class of $\{1, 1, 2\}$-TWCs with collinear node distribution.
The proposed new carbon allotrope DZQH-C$_{36}$ offers the first example of a topological TWC semimetal.
Besides realizing the new collinear TWC configuration (differing from the known triangular configuration), more importantly,
it realizes TWC semimetal in an electronic state, rather than a bosonic state as in previous studies~\cite{WangXu,C-2-WPhonos}.

The TWC states may lead to several physical consequences that can be experimentally probed.
First, the bulk Weyl nodes will dictate the existence of Fermi arc surface states. For example, for DZQH-C$_{36}$,
due to the collinear configuration of TWC, there must be ``S''-shaped Fermi arcs on the side surfaces [see Fig.~\ref{fig_surface}(b)].
This is clearly seen in Fig.~\ref{fig_surface}(a), where the surface projection of $W_2$ emits two arcs, connecting projected $W_1$ and $W_{1'}$ on its two sides. Such Fermi arc states can be directly probed using angle-resolved photoemission spectroscopy.
Second, the different chiral charges in a TWC may lead to competing effects in transport. For example, for the $\{1, 1, 2\}$-TWC in DZQH-C$_{36}$, consider transport in the $x$-$y$ plane. Carriers around the two elementary Weyl node feature weak anti-localization, whereas those around quadratic Weyl node feature weak localization. Hence, one may expect a crossover behavior when the Fermi level is shifted between these nodes. In addition, noting that in general the nodes in a TWC may reside at different energies, this could lead to a topological contribution to the circular photogalvanic effect~\cite{QuantizedCP-Moore,QNHF-FL}.

\begin{acknowledgments}
The authors thank D. L. Deng for valuable discussions. This work was supported by the National Natural Science Foundation of China (Grants No.~12304202), Hebei Natural Science Foundation (Grant No.~A2023203007), the Science Research Project of Hebei Education Department (Grant No.~BJK2024085), Quantum Science Strategic Special Project of Guangdong Province (No.~GDZX2504003), and HK PolyU start-up grant (P0057929). W.W. is supported by the Guangdong Basic and Applied Basic Research Foundation (Grant No. 2022A1515110094).
\end{acknowledgments}

\makeatletter
\let\auto@bib@innerbib\@empty
\makeatother
\begingroup
\makeatletter
\renewcommand{\bibfont}{\fontsize{9}{9.8}\selectfont\@clubpenalty\clubpenalty}
\makeatother

\endgroup

\end{document}